\documentclass[conference]{IEEEtran}

\usepackage[dvipsnames]{xcolor}
\usepackage{siunitx}

\usepackage{cite}
\usepackage{hyperref}

\usepackage{graphicx}
\usepackage[cmex10]{amsmath}
\usepackage{amssymb,mathabx}
\usepackage{dsfont}
\usepackage{tikz,pgfplots}

\usepackage{array}
\usepackage{booktabs}
\usepackage[dvipsnames]{xcolor}
\usepackage{tikz,pgfplots}
\usetikzlibrary{shadows,arrows,shapes,calc,positioning,decorations.pathreplacing,fit}
\usepackage{siunitx}
\usepgfplotslibrary{groupplots}
\pgfplotsset{compat=newest}

\DeclareMathOperator*{\argmin}{arg\,min}
\DeclareMathOperator*{\argmax}{arg\,max}

\definecolor{whinered}{RGB}{132,9,45}
\def\showpgfcircle{\tikz[baseline=-0.5ex]\node[whinered,very thick,mark size=0.5ex]{\pgfuseplotmark{o}};}
\def\showpgfsquare{\tikz[baseline=-0.6ex]\node[whinered,very thick,mark size=0.5ex]{\pgfuseplotmark{square}};}
\def\showpgfx{\tikz[baseline=-0.5ex]\node[whinered,very thick,mark size=0.6ex]{\nullfont\pgfuseplotmark{x}};}

\newcommand\blfootnote[1]{%
  \begingroup
  \renewcommand\thefootnote{}\footnote{#1}%
  \addtocounter{footnote}{-1}%
  \endgroup
}

\begin{document}

\title{Deep Learning for Communication over Dispersive Nonlinear Channels: Performance and Comparison with Classical Digital Signal Processing}

\author{\IEEEauthorblockN{Boris Karanov${}^{1,2}$, Gabriele Liga${}^{3}$, Vahid Aref${}^{2}$, Domani\c{c} Lavery${}^{1}$, Polina Bayvel${}^{1}$, Laurent Schmalen${}^{4}$\vspace*{1ex}}
\IEEEauthorblockA{${}^{1}$Optical Networks Group, University College London, WC1E 7JE London, U.K\\
${}^{2}$Nokia Bell Labs, 70435 Stuttgart, Germany\\
${}^{3}$Signal Processing Systems Group, Eindhoven University of Technology, 5600 MB Eindhoven, Netherlands\\
${}^{4}$Communications Engineering Lab, Karlsruhe Institute of Technology (KIT), 76131 Karlsruhe, Germany}
}
\IEEEspecialpapernotice{(Invited paper)}
\maketitle

\begin{abstract}
In this paper, we apply deep learning for communication over dispersive channels with power detection, as encountered in low-cost optical intensity modulation/direct detection (IM/DD) links. We consider an autoencoder based on the recently proposed sliding window bidirectional recurrent neural network (SBRNN) design to realize the transceiver for optical IM/DD communication. We show that its performance can be improved by introducing a weighted sequence estimation scheme at the receiver. Moreover, we perform bit-to-symbol mapping optimization to reduce the bit-error rate (BER) of the system. Furthermore, we carry out a detailed comparison with classical schemes based on pulse-amplitude modulation and maximum likelihood sequence detection (MLSD). Our investigation shows that for a reference 42\,Gb/s transmission, the SBRNN autoencoder achieves a BER performance comparable to MLSD, when both systems account for the same amount of memory. In contrast to MLSD, the SBRNN performance is achieved without incurring a computational complexity exponentially growing with the processed memory.

\end{abstract}

\section{Introduction}

Deep learning techniques~\cite{Goodfellow} applied\blfootnote{The work received funding from the European Union's Horizon 2020 research and innovation programme under the Marie Sk{\l}odowska-Curie project COIN (grant agreement No. 676448). G.~Liga gratefully acknowledges the European  Research  Council  (ERC)  under  the  European  Union's  Horizon 2020 research and innovation programme (grant agreement No757791)} to the design of communication systems have been subject to extensive research efforts in recent years. Often, a specific transmitter or receiver function, such as coding, modulation or equalization, is optimized using artificial neural networks (ANN) and deep learning~\cite{Khan_1,Zibar_1,Hager,Farsad_1,Houtsma}. For example, a low-complexity fiber nonlinearity compensation block was designed using ANNs in~\cite{Hager}. A deep learning-based receiver in combination with an efficient sequence estimation scheme was proposed for molecular communications~\cite{Farsad_1}. ANNs have also been considered for the equalization module in short reach optical access networks~\cite{Houtsma}. However, the approach of designing and optimizing the communication system independently on a module-by-module basis can be sub-optimal in terms of the end-to-end performance. Owing to the universal function approximator properties of the ANNs~\cite{Hornik}, it was proposed in~\cite{O'Shea_1,Doerner} to optimize the complete communication system in a single process spanning from the transmitter input to the receiver output. Such \emph{autoencoder} systems, implemented as a single deep neural network, have the potential to achieve the optimal end-to-end performance. Autoencoders recently gained popularity in communication scenarios, where the optimum pair of transmitter and receiver or optimum processing modules are not known or prohibitive due to complexity reasons. 

In low-cost optical fiber links based on intensity modulation and direct detection (IM/DD), the communication channel is nonlinear with memory, due to the joint effects of chromatic dispersion and square-law photodiode opto-electrical conversion. Such systems are particularly suitable for deep learning-based signal processing due to the absence of optimal algorithms and the presence of stringent computational cost constraints. Simple end-to-end learning systems based on feed-forward neural networks (FFNN) have been investigated for communication over such type of channels and it was verified in experiment that they can outperform conventional pulse amplitude modulation schemes with specific, ubiquitously deployed linear equalizers~\cite{Karanov_1,Chagnon}. More recently, extending the design of~\cite{Farsad_1}, an advanced autoencoder for dispersive nonlinear channels which takes into consideration the inter-symbol interference (ISI) has been proposed~\cite{Karanov_2}. The novel communication transceiver based on sliding window processing and end-to-end bidirectional recurrent neural networks (SBRNN) achieved a significant performance improvement compared to previous FFNN autoencoders by exploiting the channel memory. It was shown in~\cite{Karanov_2} that the system can outperform state-of-the-art nonlinear neural network-based equalizers using multi-symbol receiver processing~\cite{Houtsma,Lyubomirsky}, while requiring significantly fewer trainable parameters.
\begin{figure*}[t!]
\centering
\includegraphics[width=\textwidth]{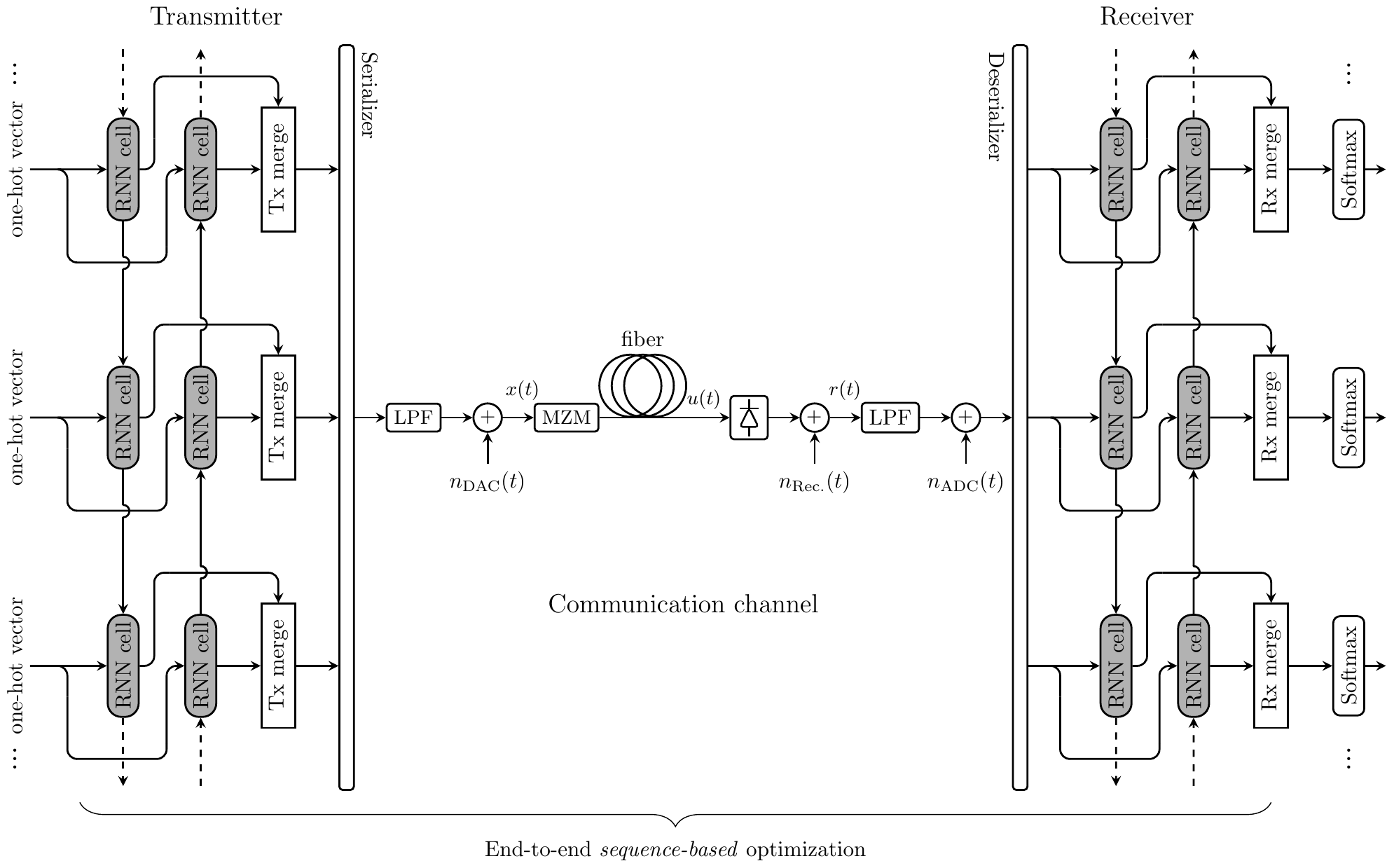}
\caption{\label{fig:Schematics} System schematic of the end-to-end SBRNN autoencoder, enabling end-to-end optimization on a sequence level via deep learning.}
\end{figure*}

In this work, we make key modifications in the SBRNN autoencoder, allowing us to increase the system reach or enhance the data rate for shorter distances. First, we generalize the sliding window estimator by introducing new weighting factors at the output of the receiver neural network. This leads to an improved bit-error rate (BER) performance compared to the previous system, where equal weights were assigned~\cite{Karanov_2}. Furthermore, instead of the previously employed ad hoc approach to bit mapping, we perform bit-to-symbol mapping optimization, thus obtaining an additional BER reduction. Additionally, we compare the end-to-end SBRNN to a benchmark system based on pulse-amplitude modulation (PAM) and classical receiver digital signal processing scheme, tailored for communication over dispersive channels, such as maximum likelihood sequence detection~(MLSD)~\cite{Forney}. Our results show that the autoencoder, when processing the same sequence time window, achieves similar BER performance to the MLSD system. However, the SBRNN scheme has a computational complexity which scales only linearly with the channel memory, contrary to the exponential growth of the MLSD scheme.

\section{Sliding Window Bidirectional Recurrent Neural Network Autoencoder}\label{sec:SBRNN} 

Following the idea of end-to-end learnable communication systems introduced in~\cite{O'Shea_1}, the complete fiber-optic link can be implemented as an end-to-end deep neural network~\cite{Karanov_1}. In particular, we employ a sequence processing scheme using a bidirectional recurrent neural network (BRNN)~\cite{Schuster}, because of the ISI from both preceding and succeeding symbols in our channel model~\cite{Karanov_2}. Figure~\ref{fig:Schematics} shows the full chain of BRNN transmitter and receiver including the communication channel.

\subsection{Bidirectional Recurrent Neural Network-Based Transceiver}\label{sec:BRNN}
\begin{figure}[t!]
\centering
\includegraphics{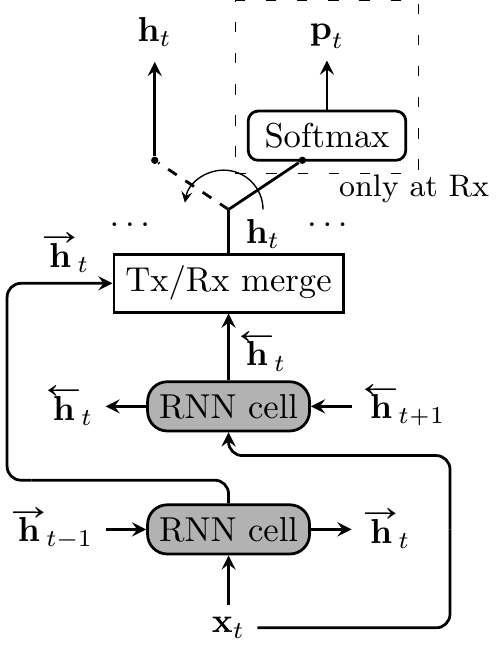}
\caption{\label{fig:BRNN_Schematics} Schematic of the bidirectional recurrent neural network processing. The final transmitter/receiver output is obtained by merging the outputs of the forward and backward passes at the same time instance. Note that \emph{softmax} activation is applied to the receiver outputs resulting in probability vectors, utilized in the sliding window estimation (see Sec.~\ref{Sliding}).}
\end{figure}

The function of the neural network-based transmitter is to encode a stream of input messages $(\ldots,m_{t-1},m_t,m_{t+1},\ldots)$,  $m_t\in\{1,\ldots, M\}$, each of which is drawn independently from an alphabet of $M$ total messages, into a sequence of transmit blocks of $n$ samples $(\ldots,\mathbf{n}_{t-1},\mathbf{n}_t,\mathbf{n}_{t+1},\ldots)$. First, we represent the messages $m_t$ as one-hot vectors $\mathbf{1}_{m,t}\in{\mathbb{R}}^{M}$ (whose elements are a ``1'' at position $m$ and zeros elsewhere) and feed them for bidirectional processing into the recurrent structure as the input $(\ldots,\mathbf{x}_{t-1},\mathbf{x}_{t},\mathbf{x}_{t+1},\ldots)$. At the receiver we apply similar bidirectional processing to the distorted samples after propagation.
The BRNN technique, identical at both transmitter and receiver, is shown schematically in Fig.~\ref{fig:BRNN_Schematics}. In the forward direction, an input $\textnormal{\textbf{x}}_{t}$ at time $t$ is processed by the recurrent cell together with the previous output $\textnormal{\textbf{h}}_{t-1}$ to produce an updated output $\textnormal{\textbf{h}}_{t}$. The procedure is performed on the full data sequence.  To adequately handle distortions arising from the succeeding symbols, the structure is also repeated in the backward direction. In this work we concentrate on a low-complexity recurrent cell based on the concatenation between the current input and the previous output. It was shown that in the framework of optical IM/DD communication, the processing capabilities of such cells are similar compared to long short-term memory structures, without the associated extra complexity~\cite{Karanov_2}. The current cell output is given by
\begin{equation*}
\mathbf{h}_{t}=\alpha_{\textnormal{Tx/Rx}}\left(\mathbf{W}\begin{pmatrix} \mathbf{x}_t^T & \mathbf{h}_{t-1}^T\end{pmatrix}^T +\mathbf{b}\right),
\end{equation*}
where ${}^{T}$ denotes the matrix transpose, $\mathbf{W}_{}\in{\mathbb{R}}^{n \times (M+n)}$ and $\mathbf{b}_{}\in{\mathbb{R}}^{n}$ (transmitter) or $\mathbf{W}_{}\in{\mathbb{R}}^{2M \times (n+2M)}$ and $\mathbf{b}_{}\in{\mathbb{R}}^{2M}$ (receiver) are the weight matrix and bias vector, respectively, and $\alpha_{\textnormal{Tx/Rx}}$ is the utilized activation function in the transmitter/receiver BRNN.

At the transmitter, the input is the one-hot vector representation of the message and the outputs of the RNN cells at the same time instance in both directions are averaged in the \emph{Tx merge} stage. Thus, at a time instant $t$, the transmitter output becomes
$\mathbf{h}_{t} = \frac{1}{2}\left({\overrightarrow{\mathbf{h}}}_{t}+{\overleftarrow{\mathbf{h}}}_{t}\right)$,
with ${\overrightarrow{\mathbf{h}}}_{t}$ and ${\overleftarrow{\mathbf{h}}}_{t}$ denoting the outputs of the RNN cells in the forward and the backward directions, respectively.
For practical purposes, we employ a transmitter activation function which limits the BRNN outputs to the $[0;\pi/4]$ range. For this reason, we use the clipping activation~\cite{Karanov_1} expressed as
\begin{equation*}
\alpha_{\text{Tx}}(\mathbf{x})=\alpha_{\tiny{\text{ReLU}}}\left(\mathbf{x}\right)-\alpha_{\tiny{\text{ReLU}}}\left(\mathbf{x}-\frac{\pi}{4}\right),
\end{equation*}
where $\mathbf{y}=\alpha_{\tiny{\text{ReLU}}}(\mathbf{x})$ is the element-wise ReLU function, i.e., $y_{i}=\max(0,x_i)$~\cite{Goodfellow}. Note that the forward/backward processing introduces extra latency in the order of a full data sequence. By termination of the sequences, we can limit this latency to a practically manageable amount (which can be in the range of thousands of symbols).

In contrast, at the receiver, we concatenate the outputs of the two RNN cells in the \emph{Rx merge} block. At time $t$, the output of the BRNN is then expressed by $\mathbf{h}_{t} = \begin{pmatrix} \overrightarrow{\mathbf{h}}_t^T & \overleftarrow{\mathbf{h}}_{t}^T\end{pmatrix}^T$. We employ pure ReLU activation, i.e., $\alpha_{\text{Rx}} = \alpha_{\tiny{\text{ReLU}}}(\mathbf{x})$. Additionally we apply a final \emph{softmax} layer after the ReLU nodes, resulting in probability vectors at the output \mbox{$\mathbf{p}_t = \mathop{\textrm{softmax}}(\mathbf{W}_{\text{softmax}}\mathbf{h}_t + \mathbf{b}_{\text{softmax}})$} with $\mathbf{p}_t\in{\mathbb{R}}^{M}$, $\mathbf{W}_{\text{softmax}}\in{\mathbb{R}}^{M\times 4M}$ and $\mathbf{b}_{\text{softmax}}\in{\mathbb{R}}^{M}$. The \emph{softmax} function is applied element-wise as $\mathbf{y}= \mathop{\textrm{softmax}}(\mathbf{x})$ with $y_{i}=\exp(x_i)/\left(\sum_{j}\exp(x_j)\right)$.

\subsection{Communication Channel}\label{sec:Channel}
Short reach optical communications are found in many data center, metro and access network scenarios and crucially rely on the simplicity and cost-effectiveness of IM/DD~\cite{Chagnon_2}. IM/DD links are also considered as prime candidates for \emph{fiber to the x} (FTTX) systems, realized for instance via passive optical networks (PONs). Such optical communication links are mainly characterized by the presence of fiber chromatic dispersion~\cite{Agrawal}, which introduces inter-symbol interference (ISI), and nonlinear (square-law) photodiode (PD) detection. Meeting the increasing data rate demands for such systems becomes quite a challenging task because of the limitations imposed by these impairments. The joint effects of ISI and square-law detection render the communication channel nonlinear with memory and one for which the optimal signal processing algorithms at transmitter and receiver are currently absent, to the best of our knowledge.

In the framework of deep learning-based autoencoder design, the communication channel is considered part of the neural network system to facilitate end-to-end training. In our work we use a detailed IM/DD channel model, i.e. an optically un-amplified link, which includes low-pass filtering (LPF) to account for practical hardware limitations at transmitter and receiver, quantization noise from the digital-to-analog (DAC) and analog-to-digital (ADC) converters, Mach-Zehnder modulator (MZM), a photodiode to perform the square-law opto-electrical conversion, electrical amplification noise and optical transmission, modeled by the attenuating and dispersive properties of the fiber. The signal that enters the receiver section of the autoencoder after channel propagation can be expressed as (neglecting the low-pass filtering at the receiver for ease of exposition)
\begin{equation*}
r(t)= |\hat{h}\{\hat{g}\{x(t)\ + n_{\text{DAC}}(t)\}\}|^{2} + n_{\text{Rec.}}(t) + n_{\text{ADC}}(t),
\end{equation*}
where $x(t)$ is the low-pass filtered transmit signal, $n_{\text{DAC/ADC}}(t)$ is additive, uniformly distributed quantization noise from the DAC/ADC, $\hat{g}\{\cdot\}$ is an operator describing the effect of the electrical field transfer function of the modulator, $\hat{h}\{\cdot\}$ describes the effects of chromatic dispersion, $n_{\text{Rec.}}(t)$ is additive Gaussian noise arising from the electrical amplification circuit at the receiver. For more details on the mathematical modeling of the channel components we refer the interested reader to~\cite[Sec.~III-B]{Karanov_1} and~\cite[Sec.~2.1]{Karanov_2}.

\section{Autoencoder Training \& Sequence Estimation}\label{Train_estimate}
In this section, we present a summary of the training procedure and explain the sequence estimation scheme in which the trained transceiver is employed. We also give a detailed description of the weighting optimization that we performed in the sliding window algorithm as well as the bit-to-symbol mapping optimization.

\subsection{Training}\label{Training}
We perform the system training in a supervised manner by using a set of labeled data. The set of transmitter and receiver BRNN parameters (denoted here by $\boldsymbol{\theta}$) is iteratively updated via stochastic gradient descent (SGD) aimed at minimizing the average loss $\underline{\mathcal{L}}(\boldsymbol{\theta})$ over a mini-batch $\underline{\mathcal{S}}$ from the training set, given by
\begin{equation}
\label{eq:BatchLoss}
\underline{\mathcal{L}}(\boldsymbol{\theta})=\frac{1}{|\underline{\mathcal{S}}|}\sum\limits_{\mathbf{1}_{m,t}\in{\underline{\mathcal{S}}}}\ell(\mathbf{1}_{m,t}, f_{\textrm{BRNN},m,t}(\ldots,\mathbf{1}_{m,t},\ldots)),
\end{equation} 
where $f_{\textrm{BRNN},m,t}(\ldots,\mathbf{1}_{m,t},\ldots)$ is the output of the end-to-end BRNN corresponding the one-hot vector $\mathbf{1}_{m,t}$ input to the transmitter at the same position and $\ell(\mathbf{x},\mathbf{y})$ denotes the loss function. In this work, we use the cross entropy as a loss function, which is defined as $\ell(\mathbf{x},\mathbf{y})=-\sum\limits_{i}x_i\log(y_i)$.
The SGD is implemented using the Adam algorithm~\cite{Kingma}. In the following we provide a detailed step-by-step description of the training procedure.

First, a set of $Z = 250$ different sequences of $T_{\textnormal{train}}=10^{6}$ random input messages $m_{i,j}$ is generated, with $i\in\{1,\ldots,Z\}$ and $j\in\{1,\ldots,T_{\textnormal{train}}\}$. At the beginning of the training, the outputs ${\overrightarrow{\mathbf{h}}}_{t-1}$ and ${\overleftarrow{\mathbf{h}}}_{t+1}$ in the forward and backward directions of the BRNN are initialized to $\mathbf{0}$. At an optimization step $s$, the mini-batch $\underline{\mathcal{S}}$ of messages $m_{i,(s-1)V+1},\ldots,m_{i,sV}$, for $1\leq i \leq Z$ and fixed $V = 10$, is processed by the transmitter BRNN to obtain the blocks $\mathbf{h}_{i,(s-1)V+1},\ldots,\mathbf{h}_{i,sV}$. Before feeding them into the communication channel, these blocks are transformed into a sequence  $\mathbf{h}_{1,(s-1)V+1},\ldots,\allowbreak\mathbf{h}_{1,sV},\mathbf{h}_{2,(s-1)V+1},\ldots,\mathbf{h}_{2,sV},\ldots,\mathbf{h}_{B,(s-1)V+1},\ldots,\mathbf{h}_{B,sV}$. At the input of the receiver, this transformation is reversed and the received blocks $\mathbf{y}_{i,(s-1)V+1},\ldots,\mathbf{y}_{i,sV}$ are applied to the BRNN, obtaining output probability vectors $\mathbf{p}_{i,(s-1)V+1},\ldots,\mathbf{p}_{i,sV}$. Then, in accordance with~\eqref{eq:BatchLoss}, the cross entropy loss between inputs and outputs is averaged over the whole mini-batch and a single iteration of the optimization algorithm is performed. Every 100 steps of the optimization, we re-initialized the outputs ${\overrightarrow{\mathbf{h}}}_{t-1}$ and ${\overleftarrow{\mathbf{h}}}_{t+1}$ in the forward and backward passes of the BRNN to $\mathbf{0}$ in an attempt to avoid local minima. Note that using validation data we confirmed that the convergence of the loss during training was achieved within 100 000 iterations, the maximum number of iterations that we allowed. 

After training of the autoencoder, we employ it in the sliding window sequence estimation algorithm proposed in~\cite{Farsad_1}. It is important to mention that for the training set a Mersenne twister was used as a random number generator. To ensure that during training we do not learn parts or construction rules of the pseudo-random sequence~\cite{Eriksson} and that training and testing datasets originate from different sources we used a Tausworthe~\cite{Lee} random number generator to generate an independent testing set of data using different 250 sequences of 10000 randomly chosen messages.

\subsection{Sliding Window Sequence Estimation Algorithm}\label{Sliding}
\begin{figure}[t!]
\centering
\includegraphics{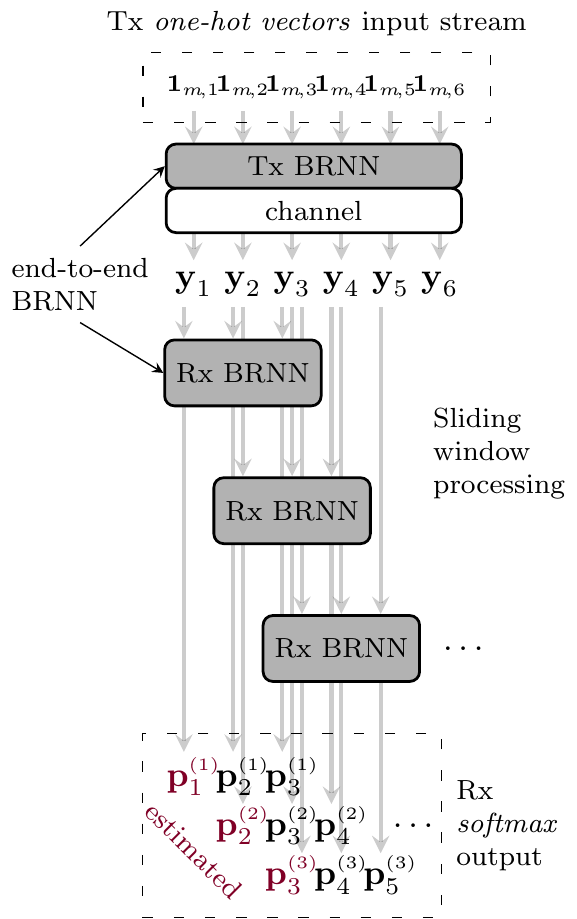}
\caption{\label{fig:SBRNN_Schematics} Schematic of the sliding window sequence estimation technique in which the BRNN transceiver is operated for $W = 3$.}
\end{figure}

Figure~\ref{fig:SBRNN_Schematics} shows a basic schematic of the sliding window sequence estimation algorithm, where the autoencoder is represented by the blocks \emph{Tx BRNN}, \emph{channel} and \emph{Rx BRNN}. For a given sequence of $T+W-1$ test messages, the transmitter BRNN encodes the full stream of input one-hot vectors $\mathbf{1}_{m,1}, \ldots, \mathbf{1}_{m,T+W-1}$. The obtained waveform is then subject to the channel, yielding the sequence of received blocks of samples $\mathbf{y}_1, \ldots, \mathbf{y}_{T+W-1}$. At a time $t$, the receiver BRNN processes the window of $W$ blocks $\mathbf{y}_t, \ldots, \mathbf{y}_{t+W-1}$, transforming them into $W$ probability vectors $\mathbf{p}_{t}^{(t)},\ldots, \mathbf{p}_{t+W-1}^{(t)}$ via its final \emph{softmax} layer. Then it slides one time slot ahead to process the blocks  $\mathbf{y}_{t+1}, \ldots, \mathbf{y}_{t+W}$. The final output probability vectors for the first $W-1$ blocks are given by
\begin{equation*}
\mathbf{p}_{i}=\frac{1}{i}\sum\limits_{k=0}^{i-1}\mathbf{p}_{i}^{(i-k)},\quad i=1, \ldots W-1.
\end{equation*}
The final probability vectors for the remaining $\mathbf{y}_{W}, \ldots, \mathbf{y}_{T}$ blocks in the received sequence are obtained as
\begin{equation}
\label{eq:Sliding_window_2}
\mathbf{p}_{i}=\sum\limits_{k=0}^{W-1}a^{(k)}\mathbf{p}_{i}^{(i-k)},\quad i=W, \ldots, T,
\end{equation}
where $a^{(q)}\geq0, q=0,\ldots, W-1$ and $\sum_{q=0}^{W-1}a^{(q)}=1$ are the weighting coefficients for the \emph{softmax} probability output of the receiver BRNN. Equal weights $a^{(q)}=\frac{1}{W}$ were previously assumed in both~\cite{Farsad_1} and~\cite{Karanov_2}. Note that the final $W-1$ blocks $\mathbf{y}_{T+1}, \ldots, \mathbf{y}_{T+W-1}$ from the received sequence are not fully estimated and we do not include them in the subsequent block error counting. We choose $T\gg W$ and thus there is only a negligible reduction in the data rate of the scheme. 

We now present a concrete example of the first few steps in the operation of the sliding window processor with $W=3$ based on Fig.~\ref{fig:SBRNN_Schematics}: At $t\!=\!1$, the receiver BRNN processes the blocks $\left(\mathbf{y}_1,\mathbf{y}_2,\mathbf{y}_3\right)$ and outputs probability vectors $\left(\mathbf{p}_1^{(t=1)},\mathbf{p}_2^{(1)},\mathbf{p}_3^{(1)}\right)$. The receiver has generated all estimates for the received block $\mathbf{y}_1$ and we have $\mathbf{p}_1^{}=\mathbf{p}_1^{(1)}$. Next, the receiver BRNN shifts a single slot to process $\left(\mathbf{y}_2,\mathbf{y}_3,\mathbf{y}_4\right)$ at $t = 2$ and generates $\left(\mathbf{p}_2^{(2)},\mathbf{p}_3^{(2)},\mathbf{p}_4^{(2)}\right)$. We have gathered all information about $\mathbf{y}^{(2)}$ and compute $\mathbf{p}_2^{}=\frac{1}{2}\left(\mathbf{p}_2^{(1)}+\mathbf{p}_2^{(2)}\right)$ as a final estimate. Similarly, after $t = 3$, a final probability vector $\mathbf{p}_3^{}=a^{(0)}\mathbf{p}_3^{(3)}+a^{(1)}\mathbf{p}_3^{(2)}+a^{(2)}\mathbf{p}_3^{(1)}$ is computed for $\mathbf{y}_3$. The sliding window processing carries on for the remainder of the blocks.

In this work, we optimize the \emph{weight coefficients} $a^{(q)}$ in order to improve the overall error rate performance. We perform the optimization offline by picking a representative test sequence of length $T+W-1$ for which we collect all corresponding BRNN output probability vectors $\mathbf{p}_{i}^{(i-k)}$, with $i=W,\ldots T$ and $k=0,\ldots, W-1$. We find the best set of coefficients $\mathbf{a}=\begin{pmatrix}a^{(0)} &\ldots & a^{(W-1)}\end{pmatrix}$ by minimizing the average cross entropy 
\begin{equation*}
\bar{c}(\mathbf{a}) := -\frac{1}{T-W+1}\sum_{i=W}^T \mathbf{1}_{m,i}^T\log\left(\sum_{q=0}^{W-1}a^{(q)}\mathbf{p}_{i}^{(i-q)}\right)
\end{equation*}
between the input one-hot vectors $\mathbf{1}_{m,W}, \ldots, \mathbf{1}_{m,T}$ and the estimated final output probability vectors $\mathbf{p}_{W}, \ldots, \mathbf{p}_{T}$, using the constrained optimization problem \begin{equation*}
    \mathbf{a}_{\textrm{opt}}=\argmin_{\mathbf{a}} \bar{c}(\mathbf{a}) \quad \textrm{s.t.} \quad a^{(q)}\geq 0,\  \sum_{q=0}^{W-1}a^{(q)}=1
\end{equation*}
where $q = 0,\ldots W-1$ and $\bar{c}$ is the average cross entropy between the input one-hot vectors and the estimated final output probability vectors.\footnote{The optimization problem is convex and can easily be solved numerically.} We apply the optimized coefficients in the estimation for all sequences in the testing set. 

\begin{figure}[t!]
\centering
\begin{tabular}{c}
\includegraphics[width=\columnwidth]{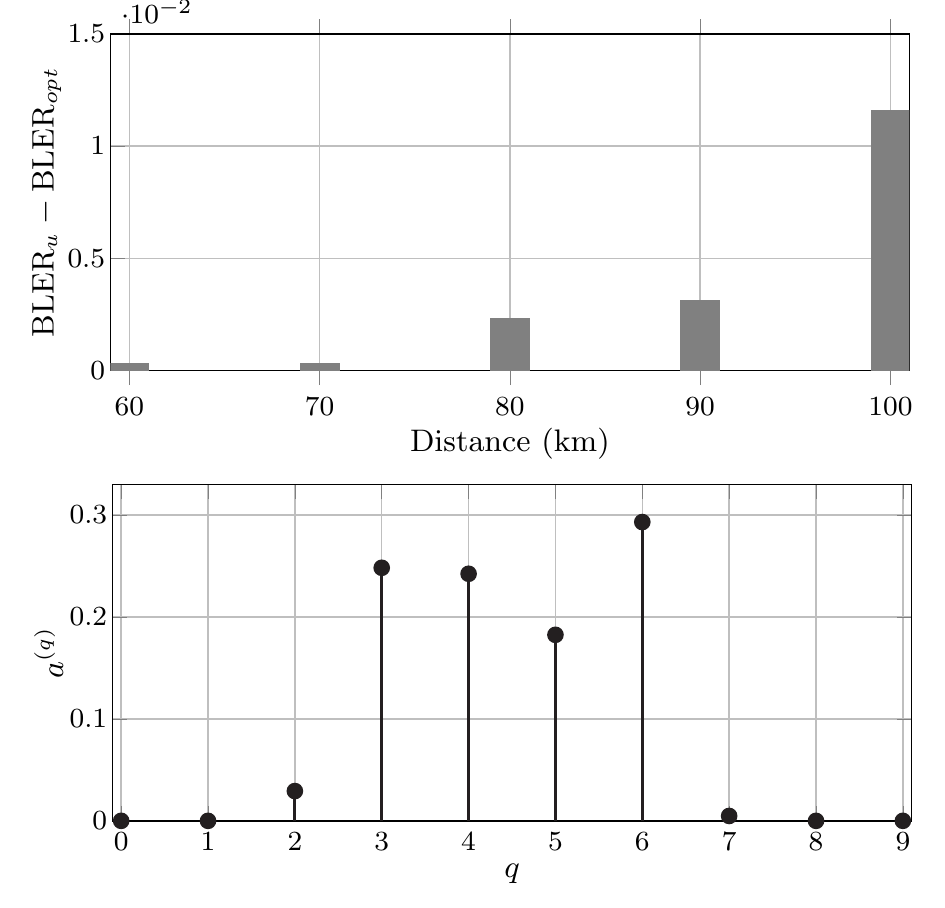}
\end{tabular}
\caption{\label{fig:coeff_opt} Top: BLER difference as a function of distance for the two approaches for final probability vector estimation in the sliding window algorithm ($W = 10$). $\textnormal{BLER}_u$ is obtained assuming $a^{(q)}=\frac{1}{W}, q=0,\ldots, W-1$ in \eqref{eq:Sliding_window_2}, while for $\textnormal{BLER}_{opt}$ the coefficients $a^{(q)}$ are optimized. Bottom: $a^{(q)}$ assignments after optimization for the system at 100\,km.}
\end{figure}

After the algorithm estimates the final probability vector for a given received block, we can perform decision on the transmitted message. We count a block error when $\argmax(\mathbf{1}_{m,i})\neq \argmax(\mathbf{p}_i)$. The block error rate (BLER) for the transmitted sequence is given by
\begin{equation*}
\text{BLER}=\frac{1}{|T_e|}\sum\limits_{i\in{T_e}}{\mathbb{I}}_{\left\{\argmax(\mathbf{1}_{m,i})\neq\argmax(\mathbf{p}_{i})\right\}},
\end{equation*}
where $|T_e|$ is the number of fully estimated messages in the test sequence and $\mathbb{I}_{\{\cdot\}}$ denotes the indicator function, equal to 1 when the argument is satisfied and 0 otherwise.

Figure~\ref{fig:coeff_opt}~(top) shows the improvement in BLER at different distances for $W=10$ after optimization of the coefficients $a^{(q)}$ in~\eqref{eq:Sliding_window_2}. As a reference we used the BLER performances when equal weights $a^{(q)}=\frac{1}{W}, q=0,\ldots, W-1$ are assigned. Note that the set of coefficients is optimized separately for each distance. We can see that the benefit from weight optimization becomes more pronounced as the distance increases. The $a^{(q)}$ assignments after optimization at 100\,km are depicted in Fig.~\ref{fig:coeff_opt}~(bottom). Larger weights are assigned to the middle indices, indicating that a received block is estimated with a higher accuracy by the receiver when relatively equal amount of pre- and post-cursor interference is captured.

\subsection{Bit Labeling Optimization}\label{sec:bitmapping}
The input and output of the ANN are non-binary messages $m\in\{1,\ldots, M\}$ and $\hat{m}$. In order to guarantee the low bit error rates (BERs) in the range $10^{-12}$ to $10^{-15}$, we require forward error correction (FEC).  For complexity reasons, FEC schemes in optical communications are usually binary~\cite{SchmalenALW} and often based on hard-decision decoding (HDD), in particular in IM/DD applications. An overview of HDD  decoding schemes and their decoding capabilities is given in~\cite{Agrell}. To convert between the ANN messages and the FEC encoder/decoder output/input, we need a \emph{bit labeling} function $\varphi: \{1,\ldots, M\}\to \mathbb{F}_2^{B}$, which maps a message $m$ to a binary vector $\mathbf{b} = \begin{pmatrix}
b_1,\ldots, b_B
\end{pmatrix}$, $b_i\in\mathbb{F}_2 = \{0,1\}$ of length $B \geq \lceil\log_2(M)\rceil$. Usually, we select $M$ and $B$ such that $M=2^B$.
\begin{figure*}[t!]\centering
\includegraphics[width=\textwidth]{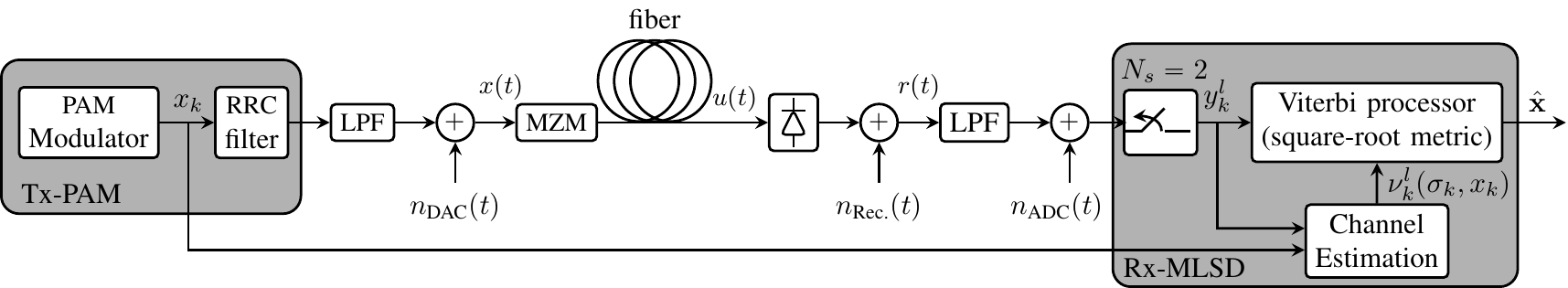}
\caption{Schematic diagram of the system used to evaluate the performance of the MLSD receiver in IM/DD optical links.}
\label{fig:MLSD_system}
\end{figure*}
Finding a bit labeling that minimizes the bit error rate is an NP-hard task that is usually solved using combinatorial optimization, e.g., a bit switching algorithm~\cite{Zeger}. Here, we use the \emph{Tabu search} algorithm~\cite{Glover} with a Tabu list of size 256. We start with a random bit labeling and, using the outcome of the validation run, namely the estimated probabilities $\hat{P}(\hat{m}|m)$, we use the expected error rate as cost function. For a given bit labeling, the Tabu search tries all possible combinations of two elements and computes the resulting expected BER. We select the combination that leads to the lowest expected BER and is not in the Tabu list. The Tabu list is then updated with this new assignment in a first-in/first-out fashion. After a pre-defined number of iterations, the overall best assignment is selected. We compare the resulting bit mapping with the trivial BER lower bound $\text{BER} > \hat{P}(\hat{m}\neq m|m)/B$, assuming that each symbol error yields exactly a single bit error.
 
Alternatively, \cite{JonesGMI} suggests to modify the autoencoder to encode a set of $B$ bits into a set of $B$ decoded bits. In this case, the loss function minimizes the average bit error rate. This is a viable alternative, however, we have found that in some circumstances, the training can get stuck in a local minimum, especially when the channel input is heavily constrained. In this work, we optimize the bit mapping which gives already performance close to the lower bound.

\section{Maximum Likelihood Sequence Detection}\label{sec:mlsd}
In order to establish a relevant benchmark for the performance of the SBRNN autoencoder, we investigate a reference $M$-PAM transmission system ($M\in\{2,4\}$) with a receiver based on maximum likelihood sequence detection~(Tx-PAM\&Rx-MLSD). For this comparison, it is worth noting that the autoencoder performs an optimization of both the transmitter and receiver designs, while in contrast, the transmitter is assumed fixed in the Tx-PAM\&Rx-MLSD scheme. However, for optical IM/DD transmission receivers based on MLSD have been widely considered \cite{Bosco2008, Poggiolini2006}, and represent a valid performance reference scheme.

The MLSD receiver assumes that the channel behaves according to a Markov process, thus facilitating the use of the Viterbi algorithm~\cite{Forney1973} which selects
\begin{equation*}
\hat{\mathbf{x}}=\argmax_{\mathbf{x}}{\hat{p}(\mathbf{y}|\mathbf{x})},
\end{equation*}
where $\mathbf{x} = (x_1,x_2,...,x_{N_{\text{sym}}})$ is the sequence of transmitted PAM symbols, $\mathbf{y}=(y^{1}_1,y^{2}_1,...,y^{N_s}_1,y^{1}_2,y^{2}_2,...,y^{N_s}_2,..., y^{N_s}_{N_{\text{sym}}})$ is the sequence of corresponding received samples. Here $N_{\text{sym}}$ and $N_s$ denote the number of PAM symbols in the transmitted sequence and number of samples per symbol, respectively. The joint distribution $\hat{p}(\mathbf{y}|\mathbf{x})$ is an approximation of the true channel likelihood $p_{\mathbf{Y}|\mathbf{X}}(\mathbf{y}|\mathbf{x})$, whose accuracy depends on the number of states and metric used in the Viterbi processor. For IM/DD systems, a more convenient option for an MLSD receiver is to operate on the distribution $p(\overline{\mathbf{y}}|\overline{\mathbf{x}})$, where $\overline{\mathbf{y}}=(\overline{y}^{1}_1,\overline{y}^{2}_1,...,\overline{y}^{N_s}_1,\overline{y}^{1}_2,\overline{y}^{2}_2,...,\overline{y}^{N_s}_2,..., \overline{y}^{N_s}_{N_{\text{sym}}})$ with $\overline{y}^{l}_{k}=\sqrt{y^{l}_{k}}$ for $l=1,...N_s$ and $k=1,2,...,N_{\text{sym}}$~\cite{Bosco2008}. Indeed, if we define the channel state as 
\begin{equation}
    \sigma_{k}\triangleq(x_{k-\frac{\mu}{2}},x_{k-\frac{\mu}{2}+1},..., x_{k-1},x_{k+1},...,x_{k+\frac{\mu}{2}}) \quad 
\label{eq:state}    
\end{equation}
where $\mu$ is the number of pre- and post-cursor symbols determining the channel memory, the distribution of $\overline{y}^{l}_{k}$ conditioned on $(\sigma_{k},x_k)$ was shown to be well approximated by a Gaussian with equal variances \cite{Bosco2008}. Under such an approximation, and the assumption of uncorrelated samples $\overline{y}^{l}_{k}$ conditional to $(\sigma_{k},x_k)$, $p(\overline{\mathbf{y}}|\overline{\mathbf{x}})$ can be suitably factorized via the channel state definition in \eqref{eq:state} for the use in a Viterbi processor.

We can thus define the branch metric of the Viterbi trellis at time $k$ as
\begin{equation}
\lambda_{k}(\sigma_{k},x_{k})=\sum_{l=1}^{N_s}\left(\sqrt{y^{l}_{k}}-\nu^{l}_{k}(\sigma_{k},x_{k})\right)^{2} \quad
\label{eq:branch_metric}
\end{equation}
where $y^{l}_{k}$ is the $l$-th out of $N_s$ samples within the $k$-th symbol period, and $\nu^{l}_{k}(\sigma_{k},x_{k})$ is given by
\begin{equation}
\nu^{l}_{k}(\sigma_{k},x_{k})=\mathbb{E}\left\{\sqrt{y^{l}_{k}(\sigma_{k},x_{k})}\right\},
\label{eq:nu}
\end{equation}
with $y^{l}_{k}(\sigma_{k},x_{k})$ indicating the value of the received sample $y^{l}_{k}$ when the pair ($\sigma_{k},x_{k}$) occurs.
The \emph{square-root metric} in \eqref{eq:branch_metric} was introduced in \cite{Poggiolini2006} and was proven to be a convenient low-complexity alternative to the histogram-based metric, yet with comparable performance \cite{Bosco2008}. At the end of each given window of $T$ symbols, the Viterbi processor makes a decision on the previous $T$ transmitted symbols by minimizing the sequence metric $\Lambda_{T}=\sum_{k=1}^{T}\lambda_{k}(\sigma_{k},x_{k})$ over all surviving state sequences $\left[\sigma_1,\sigma_2,...,\sigma_T \right]$.

\begin{figure*}[t!]
\centering
\includegraphics{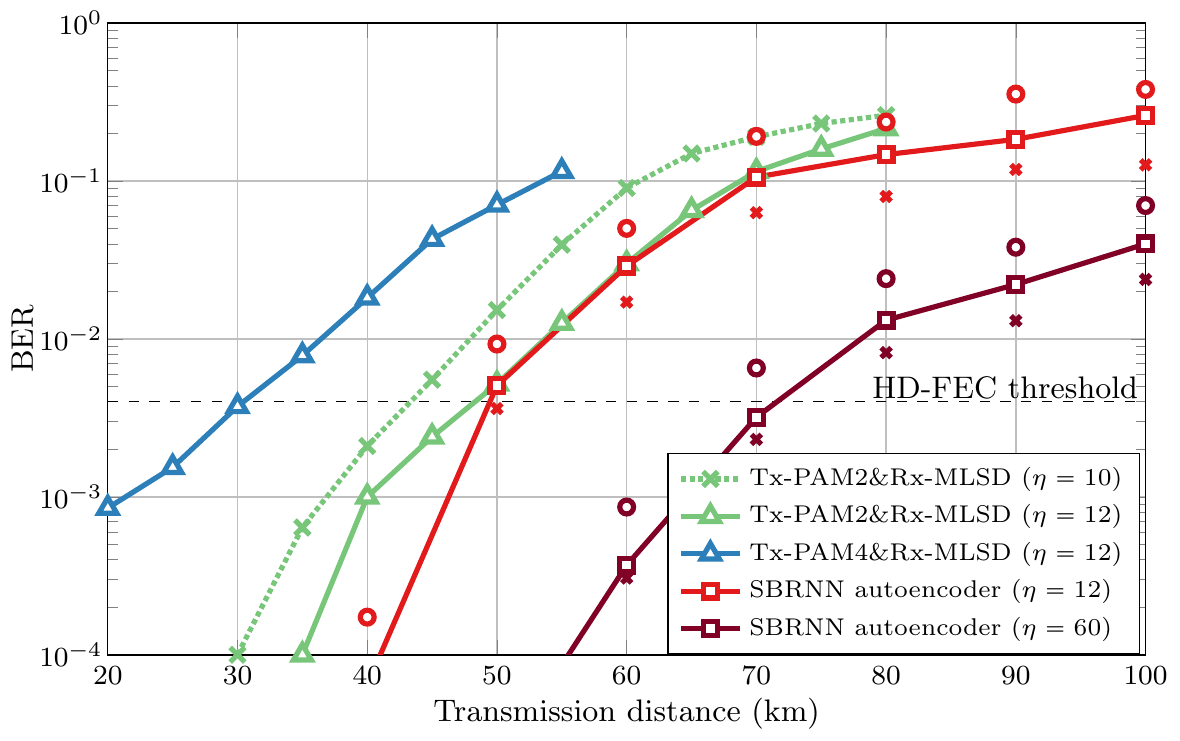}
\caption{\label{fig:Results} Bit error rate as a function of transmission distance for the 42\,Gb/s SBRNN autoencoder and $M$-PAM \& Rx MLSD systems ($M\in\{2,4\}$). In the case of MLSD $\eta = \mu\log_2(M)$, where $\mu$ represents the number of pre- and post-cursor PAM symbols defining one of $M^{\mu}$ channel states. In the case of SBRNN $\eta=W\textnormal{log}_2(M)$ is the number of bits inside the processing window. The solid SBRNN curves with square marks (\protect\showpgfsquare) include the bit-to-symbol mapping optimization described in Section~\ref{Sliding}, while circle marks (\protect\showpgfcircle) give the BER with a randomly chosen bit mapping and cross marks (\protect\showpgfx) show the lower bound on the BER (at most 1 bit error per symbol).}
\end{figure*}

The schematic diagram of the system model employing MLSD is shown in Fig.~\ref{fig:MLSD_system} (with the channel parameters taken from Table~\ref{tab:MLSD_sys_param}). At the transmitter, a Gray-labelled PAM2 ($\{0;\pi/4\}$) or  PAM4 ($\{0;\pi/12;\pi/6;\pi/4\}$) mapper was used, followed by pulse-shaping at 2 samples per symbol by a root-raised cosine (RRC) filter with a roll-off factor of 0.25. DAC/ADC rates of 84\,GSa/s and 42\,GSa/s are assumed for the PAM2 and PAM4 cases, respectively, resulting in a fixed data rate of 42\,Gb/s for both systems. A signal corresponding to a sequence of $10^5$ symbols was transmitted through the channel described in Sec.~\ref{sec:Channel}. At the receiver, the estimation of the $\nu^{l}_{k}(\sigma_{k},x_{k})$ was first done computing \eqref{eq:nu} over a sequence of $10^{7}$ symbols. The received signal, sampled at $N_s=2$ samples per symbol, is then passed directly to the Viterbi processor which performs the MLSD. 

\section{Performance and Complexity}\label{sec:Performance}
\begin{table}
 \caption{Channel Simulations Parameters}
 \label{table1:Channel_simulation_parameters}
 \centering
 \begin{tabular}{cc}
 \toprule
 Parameter & Value\\
 \midrule
 DAC/ADC rate & 84 or 42(PAM4) GSa/s \\
 LPF bandwidth & 32 GHz \\
 DAC/ADC ENOB & 6\\
 Fiber dispersion parameter & 17 ps/nm/km \\
 Fiber attenuation parameter & 0.2 dB/km \\
 Receiver noise power & $0.245$ or $0.127$(PAM4) mW  \\
 \bottomrule

 \end{tabular}\label{tab:MLSD_sys_param}
\end{table}
\begin{table}
 \caption{Autoencoder Simulations Parameters}
 \label{table2:SBRNN_simulation_parameters}
 \centering
 \begin{tabular}{cc}
 \toprule
 Parameter & Value\\
 \midrule
 \textit{M} & 64   \\
 \textit{n} & 48 \\
 Test sequence length & 10000  \\
 Processing window \textit{W} & 2 or 10 \\
 Simulation oversampling & 4 \\
 Symbol rate & 7 GSym/s \\
 Information rate & 6 bits/symbol \\
 \bottomrule

 \end{tabular}

\end{table}
In this section, we make a comparison between the BER performance of the SBRNN autoencoder and the Tx-PAM\&Rx-MLSD systems at a fixed data rate of 42 Gb/s and also discuss their computational complexity. Table~\ref{table1:Channel_simulation_parameters} lists the parameters used for the channel simulation, identical for both schemes.
\subsection{BER Performance}\label{subsec:BER_performance}
The performance of the SBRNN autoencoder is numerically evaluated using the design parameters given in Table~\ref{table2:SBRNN_simulation_parameters}. Note that weight optimization in the sliding window estimation is done as described in Sec.~\ref{Sliding}. An input sequence of messages, each from a set of $M = 64$ (6~bits), is encoded by the transmitter BRNN into a sequence of symbols (blocks) of $n = 48$ samples. We assume an oversampling factor of 4 over the 84\,GSa/s sampling rate of the DAC and, thus, simulation is performed at 336\,GSa/s. The symbol rate of the system becomes 7\,GSym/s and information is transmitted at the rate of 42\,Gb/s. In an attempt to set up a fair comparison between the investigated systems, we fixed a parameter $\eta$, which in the case of SBRNN, corresponds the number of information bits processed inside the sliding window $\eta = W\log_2M$. For the PAM transmitter with MLSD receiver, this number denotes the amount of bits contained within a channel state in the Viterbi algorithm $\eta=\mu\log_2M$, where $M\in\{2,4\}$ is the PAM order and $\mu$, already defined in Sec.~\ref{sec:mlsd}, denotes the number of post- and pre-cursor PAM symbols which form a Viterbi state. It is important to stress that in Tx-PAM2\&Rx-MLSD and Tx-PAM4\&Rx-MLSD, a fixed $\eta=12$ also means using the same number of Viterbi states (4096). This implies Tx-PAM4\&Rx-MLSD accounting for a decreased amount of memory compared to Tx-PAM2\&Rx-MLSD.

Figure~\ref{fig:Results} shows the BER performance of the examined systems as a function of transmission distance. We see that for $\eta = 12$, the SBRNN autoencoder and Tx-PAM2\&Rx-MLSD have comparable performance at all examined distances beyond 50\,km. Both systems outperform the Tx-PAM4\&Rx-MLSD, where the obtained BER is below the 6.7\% hard-decision forward error correction (HD-FEC) threshold up to 30\,km. This is due to the decreased sensitivity of the PAM4 signal and Viterbi memory ($\mu=6$). Moreover, the results indicate that assuming a wider processing window of $\eta = 60$ for the SBRNN leads to a significant BER improvement. 

~Note that we obtained the BER of the SBRNN autoencoder using the bit-to-symbol mapping algorithm described in Sec.\ref{sec:bitmapping}, for an additional comparison Fig.~\ref{fig:Results} also shows the BER performance of the system when the ad hoc approach of assigning the Gray code to the input $m\in\{1,\ldots, M\}$ is employed as well as the trivial case of a single block error resulting in a single bit error. We see that by performing the bit-to-symbol mapping optimization the performance of the system is improved, leading to an increase in the achievable distances below the HD-FEC threshold to 50\,km and 70\,km for the cases of $\eta = 12$ and $\eta = 60$, respectively.

\subsection{Computational Complexity}\label{subsec:Comp_complexity}
We study the computational complexity of the two systems using the \emph{floating point operations per decoded bit} ($\textnormal{FLOPS}_{\text{pdb}}$) as a common metric. In the case of the SBRNN autoencoder, we count FLOPS in matrix multiplications, bias additions and element-wise nonlinear activations in both direction of the recurrent structures as well as multiplications and additions in the final probability vector estimations. At the transmitter, the number of FLOPS needed to encode a bit is given by
\begin{equation*}
\text{FLOPS}_{\text{pdb}}^{[\text{SBRNN-TX}]}=\frac{2n(2(M+n)+1)}{\log_2M},
\end{equation*}
where $M$ and $n$ are hyper-parameters of the neural network, i.e. pre-defined design choices which do not change with the processing memory. Correspondingly, at the receiver we have
\begin{equation*}
\text{FLOPS}_{\text{pdb}}^{[\text{SBRNN-RX}]}=\frac{W(24M^2+8Mn+5M+2)}{\log_2M},
\end{equation*}
which exhibits a \emph{linear} dependence of the floating point operations on the processing window $W$. It is worth noting that often in practice some portion of the trained neurons are \emph{inactive}, thus reducing the actual amount of FLOPS. Furthermore, one-hot vector multiplications at the transmitter in principle can be substituted by \emph{embedding} lookups.

For the Viterbi processor, the amount of floating point operations scales linearly with the number of states in the trellis which is equal to $M^{\mu}$. Assuming a fully populated trellis, the Viterbi processor performs for each trellis section $M^{\mu+1}$ branch metric calculations, $M^{\mu+1}$ additions, and $M^{\mu+1}$ comparisons (other than the storage of $M^{\mu}$ sequences). As for the computation of the branch metric, a total of $3 N_s+(N_s-1)$ FLOPS are required (assuming addition, multiplication, and square-root use 1 FLOP each). Moreover, we will here account for each comparison as 1 FLOP and consider the additional set of comparisons required for the selection of the most likely sequence at the end of the Viterbi decoding negligible.
Thus, the overall number of required FLOPS per decoded bit is approximately given by 
\begin{equation}
\label{eq:FLOPS_per_bit_MLSD}
\text{FLOPS}_{\text{pdb}}^{[\text{MLSD}]}\approx \frac{9 M^{\mu+1}}{\log_2M}
\end{equation}
for $N_s = 2$.  Importantly, the expression in~\eqref{eq:FLOPS_per_bit_MLSD} indicates that the number of $\text{FLOPS}_{\text{pdb}}^{[\text{MLSD}]}$ exhibits an \emph{exponential} dependence on the processing memory of the detector. In presence of strong inter-symbol interference this may result in computationally prohibitive demands or increasingly sub-optimal performance.

\section{Conclusions}\label{sec:Conclusions}
We improved the performance of the SBRNN autoencoder by conducting an essential optimization of the bit-to-symbol mapping function using Tabu search combinatorial algorithm. Furthermore, we proposed an offline method for optimizing the weight assignments in the sliding window estimation algorithm which leads to BER reduction at longer distances. In a comparative study of performance and computational complexity with schemes based on PAM modulation and maximum likelihood sequence detection, our results indicate that for a fixed memory in the receiver algorithms, the SBRNN autoencoder achieves BER close to the scheme based on PAM2. It outperforms the PAM4 scheme, which is associated with higher sensitivity to noise in the system. Importantly, in terms of floating point operations per decoded bit, the autoencoder has linear dependence on the assumed memory, unlike the benchmark scheme where the dependence is exponential.

\ifCLASSOPTIONcaptionsoff
  \newpage
\fi

\end{document}